\documentclass[
 aps,
 pra,
 reprint,
 superscriptaddress,
 amsmath,
 amssymb,
 showpacs,
 nobibnotes,
 onecolumn
 ]{revtex4-2}
\usepackage{graphicx}
\usepackage{dcolumn}
\usepackage{bm}
\usepackage{feyn}
\usepackage{subfigure}
\usepackage{natbib}
\usepackage{float}
\usepackage{color}
\usepackage{amsmath}
\usepackage{amssymb}
\makeatletter

\newcommand{\Rmnum}[1]{\expandafter\@slowromancap\romannumeral #1@}
\makeatother
\usepackage{hyperref}
\hypersetup{hypertex=true,
	colorlinks=true,
	linkcolor=blue,
	citecolor=blue,
	anchorcolor=blue,
	urlcolor=blue}
\usepackage{mathrsfs}
\usepackage{rotating}
\newcommand{\RNum}[1]{\uppercase\expandafter{\romannumeral #1\relax}}
\newcommand{\be}{\begin{eqnarray}}
	\newcommand{\ee}{\end{eqnarray}}

\begin{document}
	

\title{Long-range coupling enabled multiband group-velocity control of topological edge states from slow light to light stopping}
	
\author{Junhao Yang}
\affiliation{School of Physics, Northwest University, Xi'an 710127, China}

\author{Jiarui Wang}
\affiliation{School of Physics, Northwest University, Xi'an 710127, China}

\author{Jingyu Liu}
\affiliation{School of Physical Sciences, Great Bay University, Dongguan 523000, China}

\author{Shirong Lin}
\email[Corresponding author: ]{shironglin@gbu.edu.cn}
\affiliation{School of Physical Sciences, Great Bay University, Dongguan 523000, China}
\affiliation{Great Bay Institute for Advanced Study, Dongguan 523000, China}

\author{Xinyuan Qi}
\email[Corresponding author: ]{qixycn@nwu.edu.cn}
\affiliation{School of Physics, Northwest University, Xi'an 710127, China}


\begin{abstract}
Topological edge states provide robust optical transport immune to disorder, yet their propagation velocity is usually constrained by the intrinsic band dispersion, limiting dynamic control of topological light transport.  We introduce long-range next-nearest-neighbor (NNN) couplings into a Harper--Hofstadter photonic lattice and establish a versatile platform for group-velocity engineering. We demonstrate that the NNN couplings play two distinct roles: the vertical coupling opens a previously closed topological band gap by lifting the degeneracy of bulk bands, while the horizontal coupling reshapes the edge-state dispersion through momentum-dependent corrections, enabling controllable topological slow-light transport. Furthermore, the band-gap Chern numbers associated with different gaps exhibit opposite signs, giving rise to topological edge states with opposite chiralities. Propagation simulations reveal robust unidirectional transport of these counter-chiral edge states with reduced group velocities. By continuously tuning the NNN coupling strength, the group velocity of topological edge modes can be reduced toward zero at specific momenta, resulting in topological light-stopping effects. These results demonstrate that long-range NNN couplings provide an effective mechanism for engineering momentum-dependent topological group velocities and offer new possibilities for robust slow-light devices, optical delay lines, and multiband integrated photonic systems.
\end{abstract}


\maketitle


	\section{Introduction}
The group velocity, defined as the propagation speed of a wave-packet envelope, is a fundamental physical quantity governing the dynamics of light pulses~\cite{yariv1984optical,born1999principles}. Precise control of the group velocity is central to numerous applications in modern photonics, including optical information processing, signal buffering, delay lines, and enhanced light--matter interactions for quantum technologies~\cite{Krauss2008,Parra:07,Krauss_2007}. Consequently, considerable efforts have been devoted to manipulating the group velocity of light in a variety of platforms, including atomic media~\cite{Liu2001}, engineered photonic structures~\cite{Baba2008}, and optomechanical systems~\cite{lin2026Tunable}. Phenomena such as electromagnetically induced transparency~\cite{Hau1999,Fleischhauer2005} and coherent population oscillations~\cite{Piredda2007,Boyd2009} have enabled dramatic reductions in group velocity in atomic media, even approaching zero group velocity~\cite{Liu2001}. However, these approaches often suffer from stringent experimental requirements~\cite{Hau1999}, narrow operational bandwidths~\cite{Liu2001}, limited compatibility with integrated photonic circuits~\cite{Cheng2023}, and restricted tunability~\cite{Khurgin2005,Chuang2023}, which hinder their implementation in scalable photonic platforms. Such limitations hinder their practical implementation in robust, scalable, and reconfigurable integrated photonic systems~\cite{SafaviNaeini2011}.

Artificial photonic microstructures, particularly periodic photonic structures, provide a powerful platform for dispersion engineering through structural design~\cite{Baba2008,lederer2008discrete}. By tailoring coupling strengths~\cite{Frandsen2006,Lin2023Contro}, lattice geometries~\cite{Mori2005}, and spatial symmetries~\cite{Hamachi2009}, the photonic band structure—and consequently the group velocity of supported modes—can be precisely engineered. These capabilities have enabled the realization of slow-light waveguides~\cite{Vlasov2005}, stopped-light cavities~\cite{ZHU2014154}, and tunable delay lines on chip-scale platforms~\cite{Katti2018DelayLine}. However, transport in conventional photonic lattices generally lacks topological protection~\cite{Vlasov2005,notomi2001} and remains susceptible to backscattering induced by disorder, fabrication imperfections, and sharp bends~\cite{miroshnichenko2005sharp}. This lack of robustness poses a major obstacle to reliable device operation in practical, non-ideal environments. Recently, topological photonics has emerged as a paradigm-shifting framework for realizing chiral edge states protected by bulk topological invariants through broken time-reversal symmetry~\cite{Haldane2008,Wang2009Nature}. Such states support robust unidirectional boundary transport that is highly resistant to disorder and structural imperfections~\cite{Poo2011,Lu2014,Ozawa2019,khanikaev2024,RuoYang2022}. Recent studies have begun to investigate group-velocity engineering within these topological edge channels~\cite{Guglielmon2019,yu2021topological,Rechtsman2013,Mann2021Broadband,Ma2016Valley,Kumar2024,Peng2025}, for example through dispersion engineering in Floquet topological insulators~\cite{Rechtsman2013} and valley photonic crystals~\cite{Ma2016Valley,Peng2025}. However, existing studies have primarily focused on group-velocity tuning of a single topological edge state~\cite{Kitagawa2010Floquet,Hafezi2013,Wang2025TESC}. Group-velocity engineering of multiple coexisting edge states, particularly the simultaneous realization of slow-light transport in edge states with opposite chiralities, remains largely unexplored and constitutes a significant open challenge in topological photonics. This limitation becomes particularly relevant in multiband topological systems, where different band gaps may host edge states with distinct propagation chiralities.

As a paradigmatic model in topological physics, the Harper--Hofstadter (HH) model provides an ideal platform for investigating the transport dynamics of multiple topological edge states associated with multiple band gaps~\cite{Ozawa2019,Harper1955,Hofstadter1976}. In photonic implementations, the effective magnetic flux gives rise to multiple topological band gaps, each supporting topologically protected chiral edge states, thereby offering a natural platform for studying multiband topological light transport~\cite{Harper2014,Gunnar2015,Wauters2018,Zhong2023}. To date, research on photonic HH systems has mainly focused on realizing synthetic magnetic fields, exploring topological phases, and developing topological lasers~\cite{Ye:22,Xue2023,DUTRA2023114338,Lin2022All,Matteo2019,Amelio2020}. By contrast, the dispersion engineering of topological edge states and the manipulation of their group velocities in HH lattices remain comparatively underexplored~\cite{Tai2017,Vega2023}. In particular, continuous tuning of the group velocity from the slow-light to the light-stopping regime for multiple edge states, especially those with opposite propagation chiralities, remains largely unexplored. A major limitation originates from the conventional HH model, which incorporates only nearest-neighbor (NN) couplings, severely restricting the tunability of the band structure and edge-state dispersion. This limitation motivates the exploration of extended coupling mechanisms. Recently, long-range next-nearest-neighbor (NNN) couplings have been introduced into HH lattices, enabling additional topologically protected edge channels and robust interface states between regions with different Chern numbers~\cite{Gyunghun2024}. However, their potential for independently engineering bulk topology and edge-state group velocity remains unexplored.

In this work, we introduce long-range reciprocal NNN couplings~\cite{Jia-Qi2023,Wu2020Nontrivial,Leykam2018,Bell:17,Pellerin2024,Wang2024Unconvent,Chuan2021} into a HH photonic lattice, establishing a long-range topological platform for engineering edge-state dispersion and group velocity. We show that the two NNN couplings play distinct physical roles. The vertical long-range coupling modifies the bulk-band hybridization and lifts the degeneracy responsible for the originally closed central gap, thereby opening an additional topological band gap. In contrast, the horizontal long-range coupling introduces a momentum-dependent modification of the edge-state dispersion through an additional second-harmonic contribution, enabling continuous tuning of the edge-state group velocity and the emergence of zero-group-velocity points at specific momenta. By evaluating the band-gap Chern numbers, we further reveal the coexistence of positive and negative topological invariants, corresponding to edge states with opposite propagation chiralities. Under fully open boundary conditions, the system supports topological edge transport in the first and third band gaps, while the second-gap edge states emerge under ribbon boundary conditions. Light-propagation simulations demonstrate that these counter-chiral edge modes maintain robust unidirectional transport and remain immune to structural defects while operating in the slow-light regime. The coexistence of slow-light edge states across multiple topological gaps enables multiband topological slow-light transport. Furthermore, by continuously tuning the horizontal long-range coupling, the local dispersion slope of the edge states can be compensated at isolated momenta, resulting in zero-group-velocity points and associated topological light-stopping effects confirmed by dynamical simulations. These results demonstrate that long-range NNN couplings provide an effective route toward independently engineering bulk topology and momentum-dependent edge-state group velocity. Moreover, such long-range interactions can be implemented through auxiliary-waveguide-mediated tunneling or Floquet-engineered photonic structures, making the proposed scheme compatible with existing integrated photonic platforms. Our work opens new possibilities for robust slow-light devices, optical delay lines, and multiband topological photonic systems.

\section{Long-range HH model and topological band engineering}
\begin{figure}[t]
	\centering
	\includegraphics[width=\linewidth]{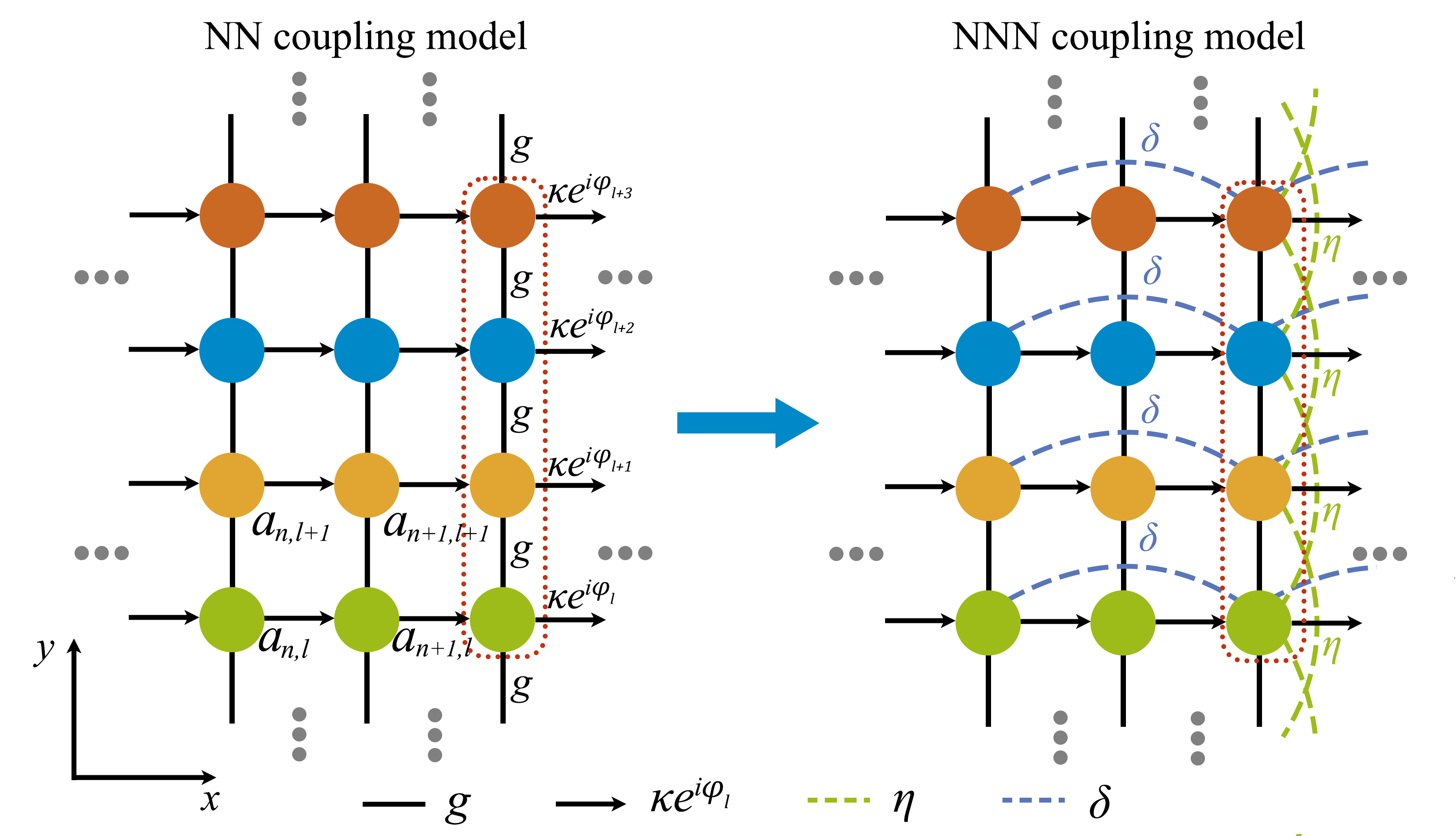}
	\caption{
		Schematic of the photonic HH lattice.
		The left and right panels show the conventional NN-coupled HH lattice and the extended HH lattice with additional NNN couplings, respectively.
		Arrows indicate the directions of nonreciprocal couplings.
		The red dashed box denotes the four-site unit cell.
		The Peierls phase is chosen as $\varphi_l=l\varphi$ with $\varphi=\pi/2$.
	}\label{fig1}
\end{figure} 
We construct an extended HH lattice model by introducing long-range NNN couplings, as illustrated in Fig.~\ref{fig1}. The left panel shows the conventional HH lattice containing only NN couplings, whereas the right panel presents the extended HH lattice with additional NNN couplings. Unlike conventional NNN couplings~\cite{Jia-Qi2023,Wu2020Nontrivial,Leykam2018,Gyunghun2024}, the long-range couplings considered here connect lattice sites separated by two lattice spacings along both the horizontal and vertical directions ($x$ and $y$ directions)~\cite{Bell:17,Pellerin2024,Wang2024Unconvent,Chuan2021}. Such couplings can also be interpreted as effective second-order coupling processes~\cite{keil2015direct}. As shown in Fig.~\ref{fig1}, $g$ and $\kappa e^{i\varphi_l}$ represent the reciprocal and nonreciprocal NN couplings, respectively, whereas $\eta$ and $\delta$ denote the reciprocal NNN couplings. A key feature of this lattice is that the phase of the nonreciprocal coupling varies with the layer index $l$ along the $y$ direction. The Peierls phase associated with the nonreciprocal coupling is defined as $\varphi=2\pi p/q$, where $p/q$ is a rational number. Consequently, the bulk eigenstates are distributed among $q$ topological energy bands characterized by nonzero Chern numbers. In the following, we focus on the representative case of $p/q=1/4$, corresponding to $\varphi=\pi/2$. Since $\varphi_l=l\varphi$, the nonreciprocal NN coupling acquires a spatially periodic phase modulation along the $y$ direction with a period of $T_{\varphi}=4$. The introduced long-range couplings do not alter this periodicity. Therefore, both the NN HH lattice and the extended NNN-coupled lattice possess a four-sublattice unit cell, as indicated by the red dashed box in Fig.~\ref{fig1}. Under periodic boundary conditions, the real-space Hamiltonian of the HH lattice with NNN couplings can be written as the sum of the NN and NNN contributions:
\begin{equation}
	\begin{aligned}
		H &= H_0+H_{N}, \\
		H_0 &= \sum_{n, l}\left(\kappa e^{i \varphi_l} a_{n+1, l}^{\dagger} a_{n, l} + g a_{n, l+1}^{\dagger} a_{n, l}+H.c.\right), \\
		H_{N} &= \sum_{n, l}\left(\delta a_{n+2, l}^{\dagger} a_{n, l}+\eta a_{n, l+2}^{\dagger} a_{n, l}+H.c.\right),
	\end{aligned} \label{eq1}
\end{equation}
where $a^\dagger$ ($a$) denotes the creation (annihilation) operator, $H_0$ describes the NN couplings, and $H_N$ describes the NNN couplings. Applying a Fourier transform to Eq.~\ref{eq1} yields the momentum-space Hamiltonian
\begin{equation}
	\begin{array}{c}
		H(k_x, k_y)=
		\begin{pmatrix}
			h_1 & g & \eta+\eta e^{-ik_y} & g e^{-ik_y} \\
			g & h_2 & g & \eta+\eta e^{-ik_y} \\
			\eta+\eta e^{ik_y} & g & h_3 & g \\
			g e^{ik_y} & \eta+\eta e^{ik_y} & g & h_4
		\end{pmatrix},
	\end{array}
	\label{eq2}
\end{equation}
where $h_\xi = 2\kappa \cos(k_x-\varphi_\xi) + 2\delta \cos(2k_x)$, with $\xi=1,2,3,4$. Since $H^\dagger = H$, the Hamiltonian is Hermitian and therefore possesses entirely real eigenvalues. Compared with the corresponding NN HH lattice, the introduction of NNN couplings breaks the chiral symmetry of the Hamiltonian, leading to modified band topology and redistribution of the topological invariants among the energy bands.

\begin{figure}[t]
	\centering
	\includegraphics[width=\linewidth]{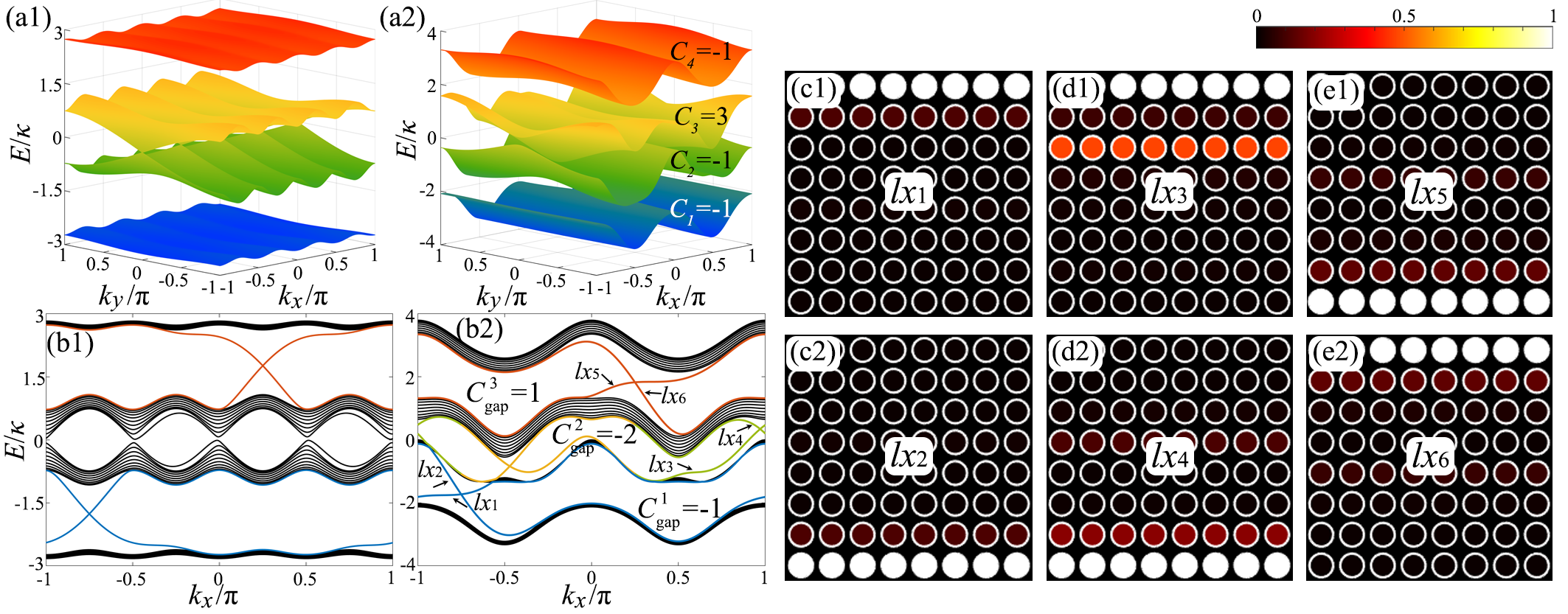}
	\caption{Band structures and edge-state spatial distributions of the HH lattices. 
		(a1,b1) Bulk and ribbon spectra of the NN HH lattice. 
		(a2,b2) Bulk and ribbon spectra of the HH lattice with NNN couplings ($N_y=40$). 
		(c1--e2) Spatial intensity distributions of representative edge states localized at the upper and lower boundaries. 
		The labels $l_{x1}$--$l_{x6}$ denote different edge-state branches. 
		For clarity, only one pair of edge states in the second band gap is shown. 
		Parameters: $\delta=0.3$, $\eta=0.3$, and $\kappa=g=1$.}\label{fig2}
\end{figure}
By solving Eq.~(\ref{eq2}), the band structures of the NN HH lattice 
($\delta=\eta=0$) and the extended NNN-coupled lattice 
($\delta=0.3$, $\eta=0.3$) are obtained, as shown in Fig.~\ref{fig2}. The comparison reveals two distinct roles of the long-range couplings. The vertical coupling $\eta$ primarily modifies the bulk band topology through additional inter-sublattice hybridization. In the NN HH lattice, the middle two bands become degenerate at high-symmetry momenta, resulting in a closed central band gap [see Fig.~\ref{fig2}(a1)]. Around these gap-closing points, the low-energy Hamiltonian can be projected onto the degenerate subspace, where the $\eta$-induced long-range coupling acts as an effective gap-opening perturbation that lifts the degeneracy and generates a finite topological gap. This mechanism is analogous to mass-term-induced gap opening in Dirac-type Hamiltonians~\cite{Hasan2010}. According to the bulk-boundary correspondence~\cite{Ozawa2019}, the newly opened gap supports additional topological edge-state branches [see Fig.~\ref{fig2}(b2)]. In contrast, the horizontal coupling $\delta$ primarily controls the momentum dependence of the edge-state dispersion. As indicated by Eq.~(\ref{eq2}), the NNN coupling $\delta$ introduces an additional second-harmonic term $2\delta\cos(2k_x)$ into the energy spectrum, which modifies the curvature and local slope of the edge-state bands. Since the group velocity is determined by $v_g=\partial E(k_x)/\partial k_x$, the $\delta$-induced dispersion correction can partially compensate the original edge-state dispersion, thereby reducing the group velocity while preserving the nonzero bulk topology and topological protection of the edge modes~\cite{Guglielmon2019,yu2021topological}. Consequently, the long-range coupling provides an additional mechanism for independently engineering the topological gap structure and edge-state transport velocity. For the NNN HH lattice, the Chern numbers of the four bands in the first Brillouin zone are calculated as ${C_\mu}=(C_1,C_2,C_3,C_4)=(-1,-1,3,-1)$~\cite{Lu2014}, where $\mu$ denotes the band index. The nonzero integer Chern numbers confirm that the system preserves its nontrivial topological character after the introduction of NNN couplings. Furthermore, analogous to the quantized Hall conductance in electronic systems~\cite{Ozawa2019}, one may define the band-gap Chern number as the sum of the Chern numbers of all occupied bands,
\begin{equation}
	C^{(r)}_{gap} = \sum_{\mu=1}^{r} C_{\mu},
	\label{eq3}
\end{equation}
where $r$ denotes the index of the band gap. According to the bulk-edge correspondence, the magnitude of the band-gap Chern number $C^{r}_{gap}$ determines the net number of chiral edge modes crossing the corresponding gap, while its sign specifies their propagation direction. The band-gap Chern numbers for the first three gaps (from lower to higher energies) are calculated from Eq.~\ref{eq3} as $(C^{1}_{gap},C^{2}_{gap},C^{3}_{gap})=(-1,-2,1)$. To verify this topological correspondence, we calculate the ribbon band structure under periodic boundary conditions along the $x$ direction and open boundary conditions along the $y$ direction, as shown in Fig.~\ref{fig2}(b). The results clearly show two edge-state branches within the second band gap, while only one branch appears in the first and third gaps. The number of edge-state branches agrees well with the absolute values of the corresponding band-gap Chern numbers, namely $|C^{1}_{gap}|$, $|C^{2}_{gap}|$, and $|C^{3}_{gap}|$. The spatial distributions further confirm that these edge states are localized at opposite boundaries of the lattice [see Figs.~\ref{fig2}(c1)--\ref{fig2}(e2)]. Due to hybridization with nearby bulk states, the edge state $l{x4}$ exhibits a more extended spatial profile. Another pair of edge states in the second band gap is also partially affected by the bulk bands, although its field distribution remains predominantly localized near the boundary. Since these states are not central to the present discussion, they are omitted from the figure for clarity. For comparison, we also calculate the ribbon spectrum of the NN HH lattice [Fig.~\ref{fig2}(b1)]. Compared with the NN HH lattice, the introduction of the long-range coupling $\delta$ significantly modifies the local slope of the edge-state dispersions in all three band gaps, resulting in reduced group velocities of the topological edge modes. Since slow-light edge states emerge across multiple topological gaps, the proposed lattice provides a platform for multiband topological slow-light transport. Here, the reduction of the edge-state group velocity is primarily governed by the horizontal NNN coupling $\delta$, while the vertical coupling $\eta$ is responsible for opening the additional topological gap.

To maintain unidirectional slow-light transport, the group velocity of the edge modes should preserve a consistent sign within the locally flattened dispersion region. This requirement imposes the condition $0.26<\delta\le0.3$: $\delta>0.26$ maintains the propagation direction of the slow-light mode in the second band gap, whereas $\delta\le0.3$ preserves the same condition for the first and third band gaps. Therefore, $\delta=0.3$ is adopted in the following calculations.

To verify that the chirality of the topological edge states is determined by the sign of the band-gap Chern number, we simulate boundary transport using the coupled-mode equation $i\dot{\bm{A}}(t)=H\bm{A}(t)$, whose formal solution is
\begin{equation}
	\bm{A}(t)=e^{-i H t}\bm{A}(0),
	\label{eq4}
\end{equation}
where $\bm{A}(t)=(A_1(t),A_2(t),\dots,A_n(t))$ denotes the field-amplitude vector at time $t$, and $\bm{A}(0)$ represents the initial excitation profile. To selectively excite the edge-state branch $l_{x(\zeta)}$, we construct a Gaussian wave packet of the form $ \bm{G}(\bm{x})=\psi_{l_{x(\zeta)}}(k_x) e^{-{(\bm{x}-x_0)^2}/{2\omega_x^2}} e^{-ik_x\bm{x}}$, where $\psi_{l_{x(\zeta)}}(k_x)$ is the eigenstate of the edge-state branch $l_{x(\zeta)}$ at momentum $k_x$ ($\zeta=1,2,\ldots,6$), $\omega_x$ denotes the wave-packet width along the $x$ direction, and $x_0$ specifies the excitation center. The initial condition is then chosen as $\bm{A}(0)=\bm{G}(\bm{x})$.

\begin{figure}[t]
	\centering
	\includegraphics[width=\linewidth]{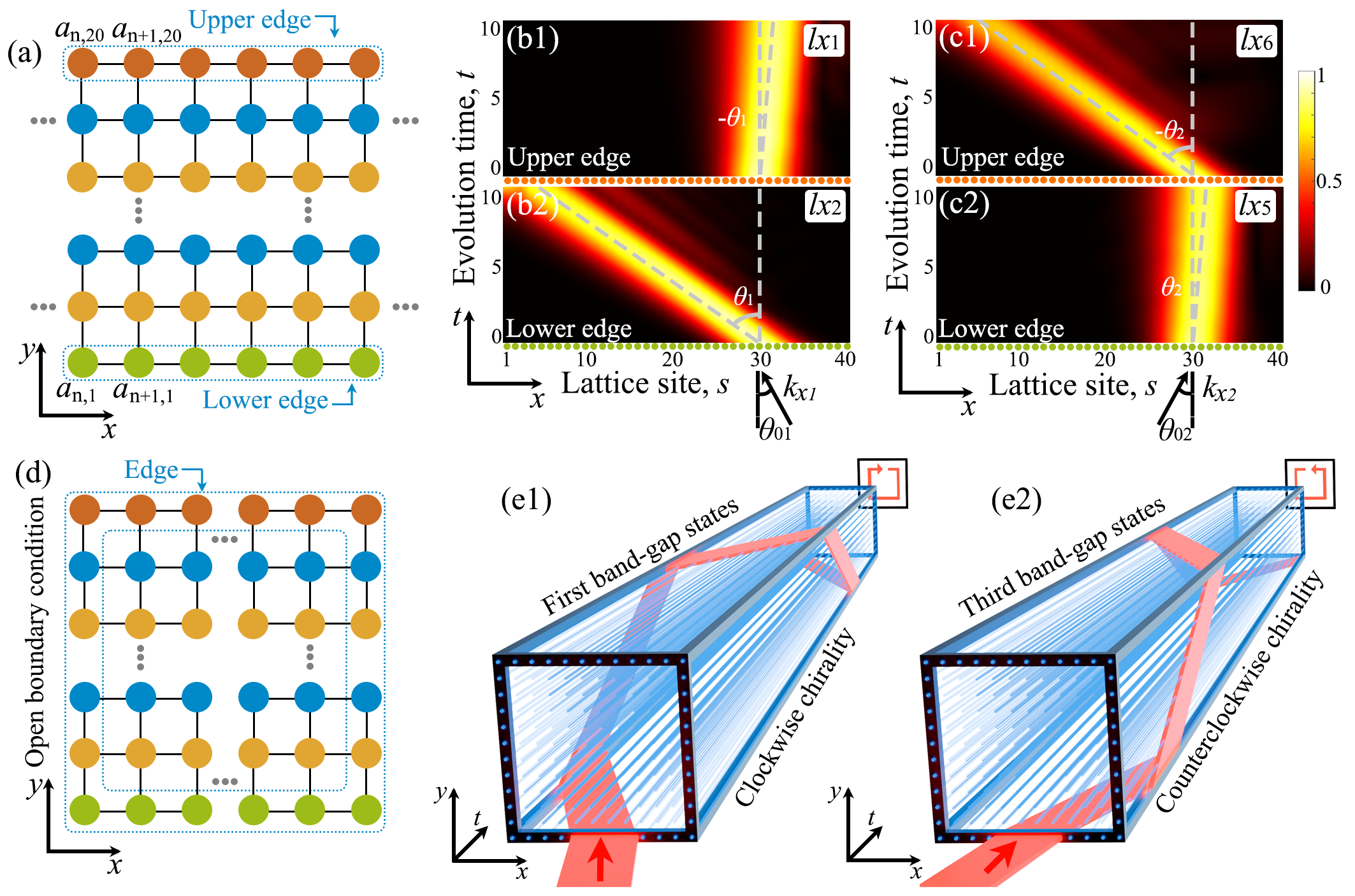}
	\caption{Chirality and boundary transport of topological edge states. (a) Momentum-space excitation regions of the upper and lower edge-state branches. 	
		(b1--c2) Propagation dynamics of different edge states in a ribbon geometry with periodic (open) boundaries along the $x$ ($y$) direction. The excitation wave vectors are chosen as $k_{x1}=-0.85\pi$ and $k_{x2}=0.35\pi$.
		(d) Schematic of the fully open boundary geometry.	(e1--e2) Schematic illustration of clockwise and counterclockwise edge-state propagation under fully open boundary conditions.}\label{fig3}
\end{figure}
As shown in Fig.~\ref{fig3}(a), the blue dashed regions indicate the excitation and propagation areas of the upper- and lower-edge states in the $x$-oriented ribbon. Selective excitation of the upper-edge state $l_{x1}$ at $k_x=-0.85\pi$ within the first band gap leads to the transport dynamics shown in Fig.~\ref{fig3}(b1). The incident and outgoing beams remain on the same side of the surface normal, exhibiting negative-refraction-like propagation~\cite{Pertsch2002,Regulating2025}. In contrast, excitation of the lower-edge state $l_{x2}$ results in the transport behavior shown in Fig.~\ref{fig3}(b2), where the outgoing beam appears on the opposite side of the surface normal, corresponding to positive-refraction-like propagation. Since the second band gap has the same sign of the band-gap Chern number as the first, it exhibits the same transport behavior and is therefore omitted for brevity.  A similar transport behavior is observed for the third-band-gap edge states at $k_x=0.35\pi$. Excitation of the upper-edge state $l_{x4}$ [Fig.~\ref{fig3}(c1)] produces negative-refraction-like propagation, whereas excitation of the lower-edge state $l_{x3}$ gives rise to positive-refraction-like propagation [Fig.~\ref{fig3}(c2)]. These results demonstrate that the upper- and lower-edge states propagate in opposite directions owing to their localization at opposite boundaries, while sharing the same chirality within the same band gap. In contrast, the chiralities associated with the first (second) and third band gaps are opposite. This behavior is fully consistent with the signs of the band-gap Chern numbers: negative (positive) band-gap Chern numbers correspond to clockwise (counterclockwise) edge transport within the slow-light regime. Based on this correspondence, we further predict the boundary propagation directions under fully open boundary conditions, as illustrated schematically in Fig.~\ref{fig3}(d). The edge states in the first band gap are expected to circulate clockwise along the system boundary [Fig.~\ref{fig3}(e1)], whereas those in the third band gap should propagate counterclockwise [Fig.~\ref{fig3}(e2)]. These predictions provide a direct connection between the band-gap Chern numbers and the boundary propagation chirality, which will be further verified through dynamical simulations below.
\begin{figure}[b]
	\centering
	\includegraphics[width=\linewidth]{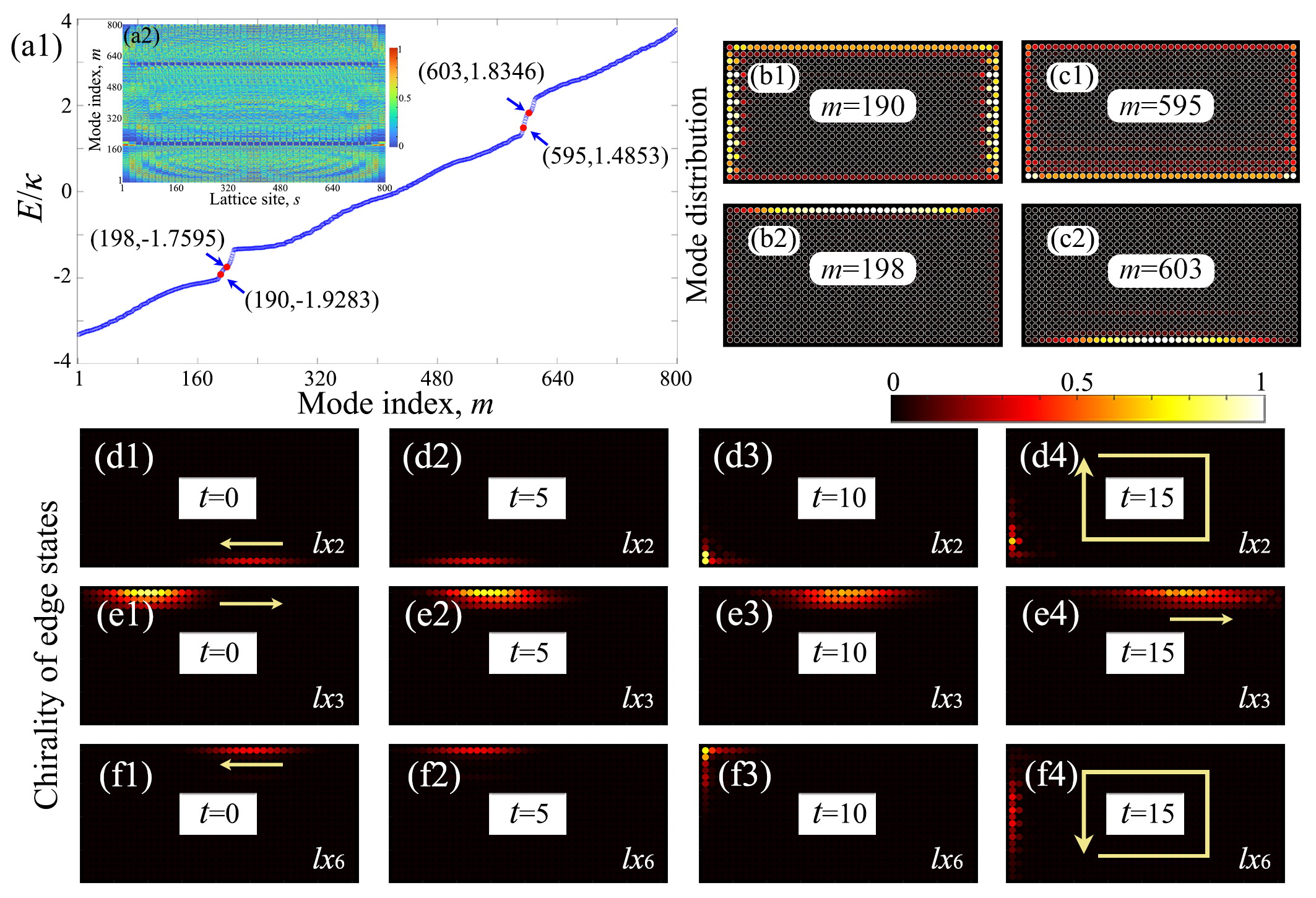}
	\caption{
		Topological edge modes and boundary transport under open boundary conditions.
		(a1) Eigenvalue spectrum of the finite HH lattice under fully open boundary conditions.
		(a2) Spatial intensity distributions of representative eigenmodes.
		(b1--b2) Representative edge modes in the first band gap (modes 190 and 198).
		(c1--c2) Representative edge modes in the third band gap (modes 595 and 603).
		(d1--d4) Clockwise propagation of the first-band-gap edge states.
		(e1--e4) Clockwise propagation of the second-band-gap edge states under the ribbon approximation.
		(f1--f4) Counterclockwise propagation of the third-band-gap edge states.
		The finite lattice contains $N_x=40$ sites along the $x$ direction and $N_y=20$ sites along the $y$ direction.
	}\label{fig4}
\end{figure}

\section{Boundary-dependent topological edge transport}
To further investigate the existence and spatial characteristics of topological edge states in finite systems, we calculate the eigenmodes of the finite-size Hamiltonian and their corresponding spatial intensity distributions under fully open boundary conditions. As shown in Figs.~\ref{fig4}(a1) and \ref{fig4}(a2), boundary-localized eigenmodes appear within the first and third band gaps, whereas no isolated edge modes are observed in the second band gap. In the first band gap, representative modes 190 and 198 exhibit a finite-size hybridized boundary state and a single-boundary localized edge state, respectively [Figs.~\ref{fig4}(b1)--\ref{fig4}(b2)], with mode 198 predominantly localized along the upper boundary. Similarly, in the third band gap, modes 595 and 603 display analogous boundary-localized characteristics [Figs.~\ref{fig4}(c1)--\ref{fig4}(c2)], with mode 603 mainly distributed along the lower boundary. These results demonstrate that the finite HH lattice supports topological edge states in the first and third band gaps under fully open boundary conditions.

The absence of isolated edge modes in the second band gap does not indicate the disappearance of topological edge states. Instead, it originates from finite-size effects in the fully open geometry. The second-gap edge states are more strongly affected by the nearby bulk bands, resulting in reduced localization and enhanced overlap between edge modes localized at different boundaries. Consequently, finite-size hybridization prevents the formation of isolated boundary eigenmodes. For sufficiently large systems, this hybridization becomes exponentially suppressed, and the local edge dynamics approaches the ribbon limit.

To verify the above prediction and further investigate the propagation dynamics of topological edge states, we solve Eq.~\ref{eq4} under the corresponding boundary configurations. When the lower-edge state in the first band gap is excited at $k_x=-0.76\pi$, the optical field propagates clockwise along the system boundary, as shown in Figs.~\ref{fig4}(d1)--\ref{fig4}(d4), consistent with the negative band-gap Chern number $C^{1}_{\rm gap}=-1$. Under the ribbon approximation, where the opposite boundaries are effectively decoupled, the upper-edge state in the second band gap at $k_x=0.75\pi$ also exhibits clockwise propagation [Figs.~\ref{fig4}(e1)--\ref{fig4}(e4)], owing to its identical band-gap Chern number sign. In contrast, excitation of the upper-edge state in the third band gap at $k_x=0.26\pi$ leads to counterclockwise propagation along the boundary [Figs.~\ref{fig4}(f1)--\ref{fig4}(f4)], corresponding to the positive band-gap Chern number $C^{3}_{\rm gap}=1$. These results confirm the direct correspondence between the sign of the band-gap Chern number and the propagation chirality of topological edge states, consistent with the prediction from bulk-edge correspondence. We next investigate the transport dynamics of the counter-chiral slow-light edge states to demonstrate their robust and controllable propagation characteristics.

\section{Topological slow light and light-stopping effect}
\begin{figure}[htbp]
	\centering
	\includegraphics[width=\linewidth]{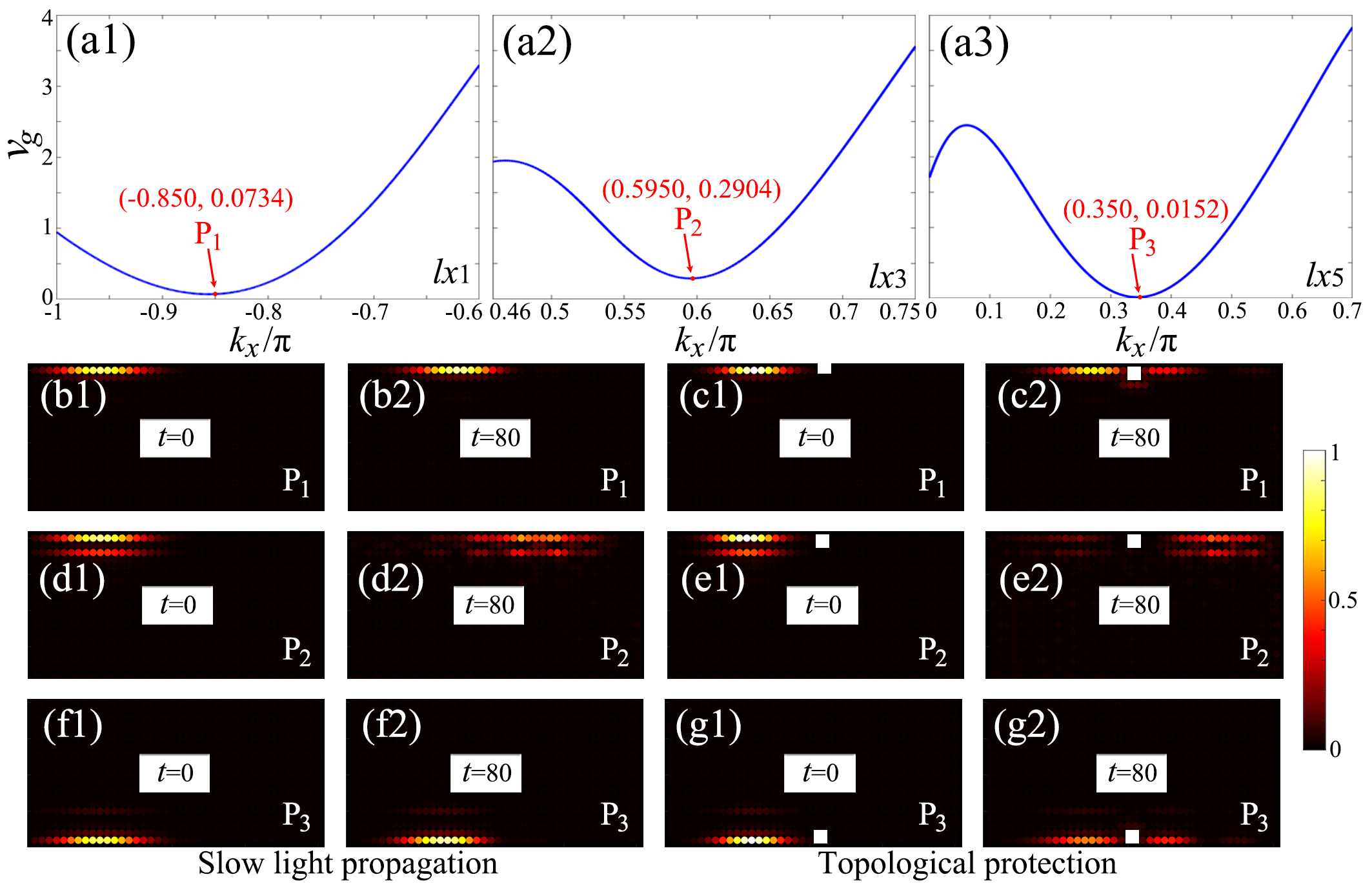}
	\caption{Multiband counter-chiral slow-light transport and robustness of topological edge states. 		(a1--a3) Group-velocity curves of slow-light edge states within the first, second, and third band gaps.
		(b1--b2), (d1--d2), and (f1--f2) Propagation dynamics of slow-light edge states in the first, second, and third band gaps, respectively.
		(c1--c2), (e1--e2), and (g1--g2) Corresponding defect-induced propagation dynamics demonstrating the robustness of slow-light edge transport in the three band gaps, respectively.}\label{fig5}
\end{figure}

To characterize the transport dynamics of topological slow-light edge states, we calculate the group velocity from the edge-state dispersion relation. The group velocity, $v_g=\partial E(k_x)/\partial k_x$, characterizes the propagation velocity of the wave-packet center for a sufficiently narrow excitation in momentum space, where $E(k_x)$ denotes the edge-state dispersion. Therefore, we calculate the group-velocity curves within the slow-light regime for the first, second, and third band gaps, as shown in Figs.~\ref{fig5}(a1--a3). The arrows indicate the operating points adopted in the subsequent propagation simulations, where the operating points are selected near the minimum positive group velocities while maintaining unidirectional propagation.

At the operating point $P_1$ in Fig.~\ref{fig5}(a1), corresponding to $k_x=-0.850\pi$, the group velocity is reduced to $v_g=0.0734$, approaching the zero-group-velocity limit while preserving directional edge transport. Consequently, the excited wave packet remains close to its excitation position even after long-distance evolution. As shown in Figs.~\ref{fig5}(b1--b2), at the evolution time $t=80$, the slow-light edge state in the first band gap remains localized near the excitation point while propagating rightward along the upper boundary, consistent with its clockwise chirality. Although the second-gap edge states disappear under fully open boundary conditions due to finite-size hybridization, they remain well-defined in the ribbon geometry and exhibit analogous slow-light transport. Choosing the operating point $P_2$ in Fig.~\ref{fig5}(a2) yields a group velocity of $v_g=0.2904$. As shown in Figs.~\ref{fig5}(d1--d2), although the propagation is faster than that in the first band gap, the wave packet still propagates substantially more slowly than the conventional edge state excited at $k_x=0.75\pi$ [see Fig.~\ref{fig4}(e4)], while maintaining rightward transport along the upper boundary. In contrast, for the third band gap, the operating point $P_3$ [ see Fig.~\ref{fig5}(a3), $k_x=0.350\pi$] corresponds to an even smaller group velocity of $v_g=0.0152$. Owing to the counterclockwise chirality, the wave packet propagates rightward along the lower boundary. At $t=80$, it remains closer to the excitation position than that in the first band gap, demonstrating extremely slow topological edge transport. To examine the robustness of the slow-light modes, defects are intentionally introduced along the propagation paths in all three band gaps. The corresponding propagation dynamics are presented in Figs.~\ref{fig5}(c1--c2), \ref{fig5}(e1--e2), and \ref{fig5}(g1--g2). In all cases, the slow-light edge states bypass the defects and continue propagating along the boundaries without noticeable backscattering, demonstrating that the edge transport remains robust against moderate structural defects without noticeable backscattering. These results establish multiband, counter-chiral, topologically protected slow-light transport in the proposed long-range-coupled HH lattice.

\begin{figure}[htbp]
	\centering
	\includegraphics[width=\linewidth]{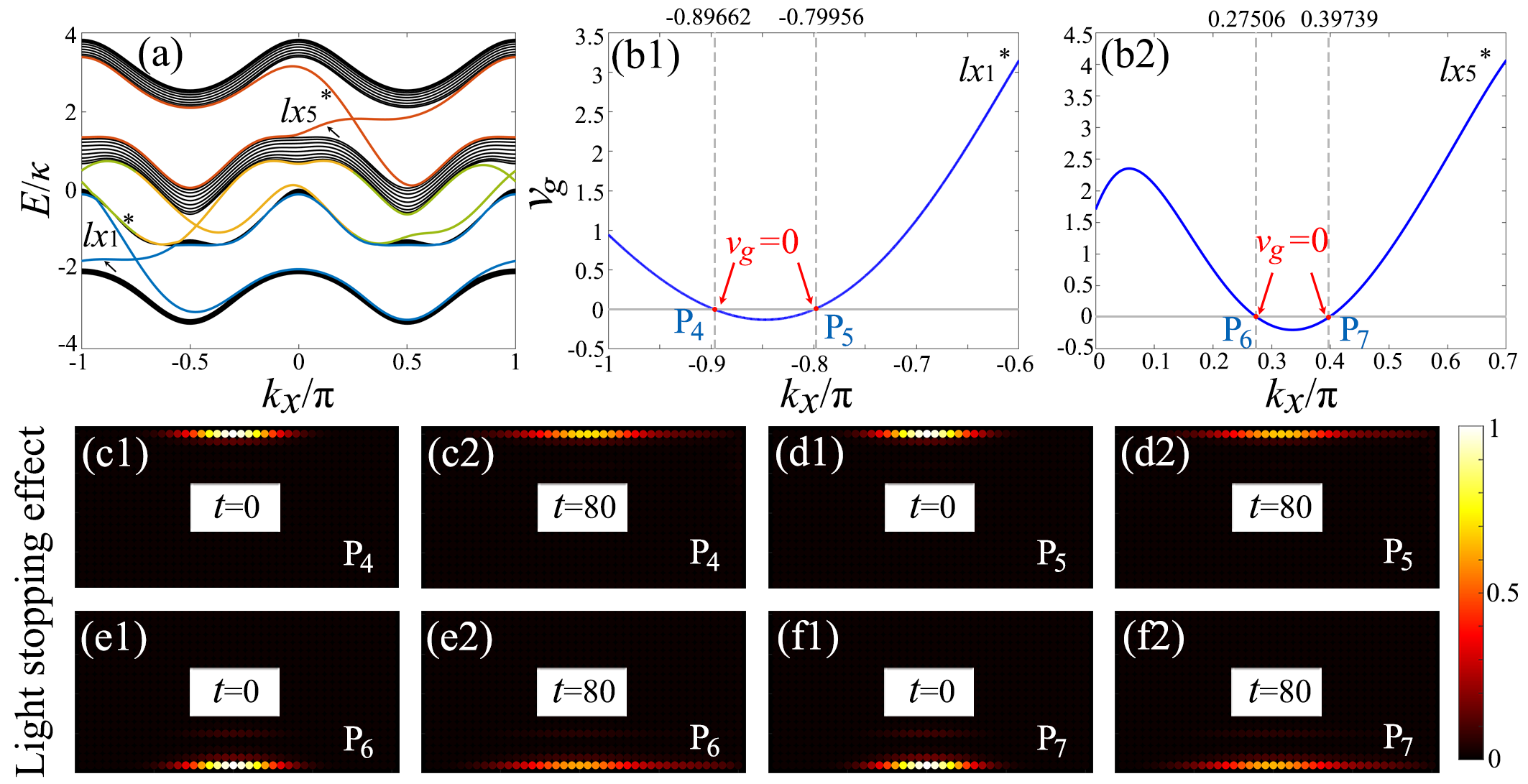}
	\caption{Topological light stopping at zero-group-velocity points. (a) Ribbon band structure for $\delta=0.32$. (b1--b2) Group-velocity curves of the first- and third-band-gap edge states in the slow-light regime.	The points $P_4$--$P_7$ indicate the momenta where the group velocity reaches zero ($v_g=0$). (c1--f2) Propagation dynamics of wave packets excited at the zero-group-velocity points, demonstrating the topological light-stopping effects.}\label{fig6}
\end{figure}

Beyond the unidirectional slow-light regime, we further explore a larger long-range coupling strength by setting $\delta=0.32$. The corresponding ribbon band structure is shown in Fig.~\ref{fig6}(a). Compared with the previous regime, the edge-state dispersions in the first and third band gaps are further reshaped, while the second-band-gap dispersion becomes more dispersive. The corresponding group-velocity curves in Figs.~\ref{fig6}(b1--b2) reveal that the group velocity is no longer strictly positive, indicating the breakdown of unidirectional slow-light transport. Meanwhile, isolated zero-group-velocity points emerge when the local slope of the edge-state dispersion vanishes. These points correspond to the compensation between the NN- and NNN-induced dispersion contributions, resulting in stationary topological edge wave packets. As indicated by the arrows in Figs.~\ref{fig6}(b1--b2), zero-group-velocity points appear at $P_4$ and $P_5$ in the first band gap and at $P_6$ and $P_7$ in the third band gap. Dynamical simulations shown in Figs.~\ref{fig6}(c1--f2) demonstrate that the excited wave packets remain localized around their initial positions with only moderate spatial broadening during evolution. These results confirm the emergence of multiple topological light-stopping effects induced by long-range dispersion engineering.

\section{Experimental feasibility and implementation strategy}

In femtosecond-laser-written photonic waveguide arrays, periodic modulation of the propagation trajectories, such as helical bending or segmented transverse displacement, provides a natural route for Floquet engineering of HH lattices~\cite{Rechtsman2013,Maczewsky2017,Sebabrata2020}. The longitudinal modulation introduces an effective synthetic gauge field along the propagation direction, generating complex Peierls phases in the NN couplings and realizing an effective Floquet Hamiltonian with HH-type topology. Building on this Floquet-engineered NN platform, detuned auxiliary waveguides can be introduced to generate effective long-range couplings through virtual tunneling processes. In this scheme, the auxiliary channels serve as intermediate states with controllable propagation-constant detuning relative to the main lattice~\cite{Szameit_2010}, mediating hopping between spatially separated sites. In the large-detuning regime, the auxiliary modes can be adiabatically eliminated~\cite{Mrejen2015}, leading to an effective second-order coupling described by perturbation theory. Specifically, the induced hopping amplitude follows the scaling relation $J_{\rm eff}\sim-\frac{J_1J_2}{\Delta}$, where $J_1$ and $J_2$ denote the couplings between the main and auxiliary waveguides, and $\Delta$ represents the propagation-constant detuning. By appropriately arranging auxiliary waveguides along both transverse directions, effective anisotropic NNN couplings can be engineered within the Floquet-averaged framework. The Floquet modulation and auxiliary-waveguide-mediated coupling provide complementary control parameters, enabling the NN Peierls phases and long-range hopping amplitudes to be engineered separately to a good approximation. The resulting photonic platform therefore realizes a practical implementation scheme combining Floquet-induced complex NN coupling and auxiliary-mediated long-range tunneling, providing an experimentally accessible route toward the proposed extended HH lattice.

\section{Conclusions}
In summary, we have introduced long-range reciprocal NNN couplings into a HH photonic lattice, establishing a topological photonic platform for separately controlling bulk topology and edge-state group velocity. We have demonstrated that the vertical long-range coupling $\eta$ modifies the bulk-band hybridization and lifts the degeneracy responsible for the originally closed central topological gap, thereby generating an additional topological band gap. In contrast, the horizontal coupling $\delta$ reshapes the edge-state dispersion through a momentum-dependent correction, providing an effective route for reducing the group velocity of topological edge modes while maintaining the topological edge-state characteristics. By calculating the band-gap Chern numbers, we reveal the coexistence of positive and negative topological invariants, corresponding to edge states with opposite propagation chiralities in different topological gaps. Under fully open boundary conditions, the system supports edge-mode transport in the first and third band gaps, whereas the second-band-gap edge states are revealed only in the ribbon geometry due to finite-size hybridization under fully open boundaries. Light-propagation simulations further demonstrate that these counter-chiral edge modes exhibit unidirectional transport and remain robust against structural defects in the slow-light regime. The presence of slow-light edge states across multiple topological gaps enables multiband topological slow-light transport. Furthermore, by tuning the horizontal long-range coupling $\delta$, the local slope of the edge-state dispersion can be continuously reduced through compensation of the intrinsic group-velocity contribution, leading to isolated zero-group-velocity points. These zero-group-velocity edge states, confirmed by dynamical simulations, provide a new mechanism for realizing topological light stopping. Moreover, the proposed long-range HH lattice can be implemented in femtosecond-laser-written photonic waveguide arrays by combining Floquet modulation for complex NN couplings with auxiliary-waveguide-mediated virtual tunneling for effective NNN interactions. Our results establish long-range NNN coupling as an effective dispersion-engineering strategy for controlling topological edge-state group velocity, offering new possibilities for robust slow-light devices, optical delay lines, and integrated multiband photonic systems.

\section{Acknowledgement}
This work was supported by the National Natural Science Foundation of China (Grants No. 12174307 and No. 12104296). We thank Prof. Ruo-Yang Zhang for fruitful discussions and helpful advice.

\nocite{*}
\bibliography{bibliography2}

@CONTROL{REVTEX42Control}

@CONTROL{apsrev42Control,author="08",editor="1",pages="0",title="0",year="1"}

@book{yariv1984optical,
  title={Optical Waves in Crystals: Propagation and Control of Laser Radiation},
  author={Yariv, A. and Yeh, P.},
  isbn={9780471091424},
  lccn={83006892},
  series={A Wiley interscience publication},
  url={https://books.google.com/books?id=jjzxAAAAMAAJ},
  year={1984},
  publisher={Wiley}
}

@book{born1999principles,
  title={Principles of Optics: Electromagnetic Theory of Propagation, Interference and Diffraction of Light},
  author={Born, M. and Wolf, E. and Bhatia, A.B.},
  isbn={9780521642224},
  lccn={98115588},
  url={https://books.google.com/books?id=nUHGpfNsGyUC},
  year={1999},
  publisher={Cambridge University Press}
}

@article{Krauss2008,
  author   = {Thomas F. Krauss},
  title    = {Why do we need slow light?},
  journal  = {Nat. Photonics},
  year     = {2008},
  volume   = {2},
  number   = {8},
  pages    = {448--450},
  month    = {aug},
  doi      = {10.1038/nphoton.2008.139}
}

@article{Parra:07,
author = {Enrique Parra and John R. Lowell},
journal = {Opt. Photon. News},
number = {11},
pages = {40--45},
publisher = {Optica Publishing Group},
title = {Toward Applications of Slow Light Technology},
volume = {18},
month = {Nov},
year = {2007},
url = {https://www.optica-opn.org/abstract.cfm?URI=opn-18-11-40},
doi = {10.1364/OPN.18.11.000040},
}

@article{Krauss_2007,
doi = {10.1088/0022-3727/40/9/S07},
url = {https://doi.org/10.1088/0022-3727/40/9/S07},
year = {2007},
month = {apr},
publisher = {},
volume = {40},
number = {9},
pages = {2666},
author = {Krauss, T F},
title = {Slow light in photonic crystal waveguides},
journal = {J. Phys. D. Appl. Phys.}
}

@article{Liu2001,
  author  = {C. Liu and Z. Dutton and C. H. Behroozi and L. V. Hau},
  title   = {Observation of coherent optical information storage in an atomic medium using halted light pulses},
  journal = {Nature},
  volume  = {409},
  pages   = {490--493},
  year    = {2001},
  doi     = {10.1038/35054017},
  url     = {https://www.nature.com/articles/35054017}
}

@article{Baba2008,
  author  = {T. Baba},
  title   = {Slow light in photonic crystals},
  journal = {Nat. Photonics},
  volume  = {2},
  pages   = {465--473},
  year    = {2008},
  doi     = {10.1038/nphoton.2008.146},
  url     = {https://doi.org/10.1038/nphoton.2008.146}
}

@article{lin2026Tunable,
  title={Tunable rotation-associated slow-to-fast light conversion via optomagnonic coupling},
  author={Liu, Jingyu and Lin, Shirong},
  journal={Phys. Rev. A},
  volume={113},
  number={5},
  pages={053514},
  year={2026},
  publisher={APS},
  doi = {10.1103/nsv9-177g},
  url = {https://link.aps.org/doi/10.1103/nsv9-177g}
}

@article{Hau1999,
  author   = {Lene Vestergaard Hau and S. E. Harris and Zachary Dutton and Cyrus H. Behroozi},
  title    = {Light speed reduction to 17 metres per second in an ultracold atomic gas},
  journal  = {Nature},
  year     = {1999},
  volume   = {397},
  number   = {6720},
  pages    = {594--598},
  month    = {feb},
  doi      = {10.1038/17561}
}

@article{Fleischhauer2005,
  author  = {M. Fleischhauer and A. Imamoglu and J. P. Marangos},
  title   = {Electromagnetically induced transparency: Optics in coherent media},
  journal = {Rev. Mod. Phys.},
  volume  = {77},
  pages   = {633--673},
  year    = {2005},
  doi     = {10.1103/RevModPhys.77.633},
  url     = {https://journals.aps.org/rmp/abstract/10.1103/RevModPhys.77.633}
}

@article{Piredda2007,
	author = {G. Piredda and R. W. Boyd},
	title = {Slow light by means of coherent population oscillations: laser linewidth effects},
	DOI= {10.2971/jeos.2007.07004},
	url= {https://doi.org/10.2971/jeos.2007.07004},
	journal = {J. Eur. Opt. Soc.-Rapid Publ.},
	year = {2007},
	volume = {2},
	pages = {07004}
}

@article{Boyd2009,
author = {Robert W. Boyd  and Daniel J. Gauthier },
title = {Controlling the Velocity of Light Pulses},
journal = {Science},
volume = {326},
number = {5956},
pages = {1074-1077},
year = {2009},
doi = {10.1126/science.1170885},
URL = {https://www.science.org/doi/abs/10.1126/science.1170885}
}

@article{Cheng2023,
  author  = {Pei Cheng and Zhongyin Xiao and Xuxian Jiang and Yulong Liu and Xianshun Cai},
  title   = {A Reconfigurable Three-Dimensional Electromagnetically Induced Transparency Metamaterial with Low Loss and Large Group Delay},
  journal = {Electronics},
  volume  = {12},
  number  = {24},
  pages   = {4930},
  year    = {2023},
  doi     = {10.3390/electronics12244930},
  url     = {https://doi.org/10.3390/electronics12244930}
}

@article{Khurgin2005,
  author = {Khurgin, J. B.},
  title = {Optical buffers based on slow light in electromagnetically induced transparent media and coupled resonator structures: comparative analysis},
  journal = {J. Opt. Soc. Am. B},
  volume = {22},
  number = {5},
  pages = {1062--1074},
  year = {2005},
  doi = {10.1364/JOSAB.22.001062},
  url = {https://doi.org/10.1364/JOSAB.22.001062}
}

@article{Chuang2023,
  author  = {You-Lin Chuang},
  title   = {Significant enhancement of group delay in electromagnetically induced transparency with a spatially partially coherent coupling field},
  journal = {Phys. Rev. A},
  volume  = {108},
  pages   = {063707},
  year    = {2023},
  doi     = {10.1103/PhysRevA.108.063707},
  url     = {https://doi.org/10.1103/PhysRevA.108.063707}
}

@article{SafaviNaeini2011,
  author  = {Safavi-Naeini, A. H. and Alegre, T. P. Mayer and Chan, J. and Eichenfield, M. and Winger, M. and Lin, Q. and Hill, J. T. and Chang, D. E. and Painter, O.},
  title   = {Electromagnetically induced transparency and slow light with optomechanics},
  journal = {Nature},
  volume  = {472},
  pages   = {69--73},
  year    = {2011},
  doi     = {10.1038/nature09933},
  url     = {https://doi.org/10.1038/nature09933}
}

@article{lederer2008discrete,
title = {Discrete solitons in optics},
journal = {Phys. Rep.},
volume = {463},
number = {1},
pages = {1-126},
year = {2008},
issn = {0370-1573},
doi = {https://doi.org/10.1016/j.physrep.2008.04.004},
url = {https://www.sciencedirect.com/science/article/pii/S0370157308001257},
author = {Falk Lederer and George I. Stegeman and Demetri N. Christodoulides and Gaetano Assanto and Moti Segev and Yaron Silberberg}
}

@article{Frandsen2006,
  author  = {Lars H. Frandsen and Andrei V. Lavrinenko and Jacob Fage-Pedersen and Peter I. Borel},
  title   = {Photonic crystal waveguides with semi-slow light and tailored dispersion properties},
  journal = {Opt. Express},
  volume  = {14},
  pages   = {9444--9450},
  year    = {2006},
  doi     = {10.1364/oe.14.009444},
  url     = {https://doi.org/10.1364/oe.14.009444}
}

@article{Lin2023Contro,
    author = {Lin, Shirong and Liang, Yao and Zhang, Jingcheng and Chen, Mu Ku and Tsai, Din Ping},
    title = {Controllable flatbands via non-Hermiticity},
    journal = {Appl. Phys. Lett.},
    volume = {123},
    number = {22},
    pages = {221103},
    year = {2023},
    month = {11},
    issn = {0003-6951},
    doi = {10.1063/5.0174456},
    url = {https://doi.org/10.1063/5.0174456}
}

@article{Mori2005,
  author  = {D. Mori and T. Baba},
  title   = {Wideband and low dispersion slow light by chirped photonic crystal waveguide},
  journal = {Opt. Express},
  volume  = {13},
  pages   = {9398--9408},
  year    = {2005},
  doi     = {10.1364/OPEX.13.009398},
  url     = {https://doi.org/10.1364/OPEX.13.009398}
}

@article{Hamachi2009,
author = {Yohei Hamachi and Shousaku Kubo and Toshihiko Baba},
journal = {Opt. Lett.},
number = {7},
pages = {1072--1074},
publisher = {Optica Publishing Group},
title = {Slow light with low dispersion and nonlinear enhancement in a lattice-shifted photonic crystal waveguide},
volume = {34},
month = {Apr},
year = {2009},
doi = {10.1364/OL.34.001072}
}

@article{Vlasov2005,
  author  = {Vlasov, Yurii A. and O'Boyle, Martin and Hamann, Hendrik F. and McNab, Sharee J.},
  title   = {Active control of slow light on a chip with photonic crystal waveguides},
  journal = {Nature},
  volume  = {438},
  pages   = {65--69},
  year    = {2005},
  doi     = {10.1038/nature04210},
  url     = {https://doi.org/10.1038/nature04210}
}

@article{ZHU2014154,
author = {Na Zhu and Yige Wang and Qingqing Ren and Li Zhu and Minmin Yuan and Guimin An},
title = {Slow light in nonlinear photonic crystal coupled-cavity waveguides},
journal = {Opt. Laser Technol.},
volume = {57},
pages = {154-158},
year = {2014},
issn = {0030-3992},
doi = {https://doi.org/10.1016/j.optlastec.2013.10.009},
url = {https://www.sciencedirect.com/science/article/pii/S0030399213003745}
}

@article{Katti2018DelayLine,
  author  = {Katti, Rohan},
  title   = {Photonic Delay Lines Based on Silicon Coupled Resonator Optical Waveguide Structures},
  journal = {Silicon},
  volume  = {10},
  pages   = {2793--2800},
  year    = {2018},
  doi     = {10.1007/s12633-018-9819-y},
  url     = {https://doi.org/10.1007/s12633-018-9819-y}
}

@article{notomi2001,
  title = {Extremely Large Group-Velocity Dispersion of Line-Defect Waveguides in Photonic Crystal Slabs},
  author = {Notomi, M. and Yamada, K. and Shinya, A. and Takahashi, J. and Takahashi, C. and Yokohama, I.},
  journal = {Phys. Rev. Lett.},
  volume = {87},
  issue = {25},
  pages = {253902},
  numpages = {4},
  year = {2001},
  month = {Nov},
  publisher = {American Physical Society},
  doi = {10.1103/PhysRevLett.87.253902},
  url = {https://link.aps.org/doi/10.1103/PhysRevLett.87.253902}
}

@article{miroshnichenko2005sharp,
  title={Sharp bends in photonic crystal waveguides as nonlinear Fano resonators},
  author={Andrey E. Miroshnichenko and Yuri S. Kivshar},
  journal={Opt. Express},
  year={2005},
  volume={13 11},
  pages={3969-76},
  url={https://api.semanticscholar.org/CorpusID:7452799}
}

@article{Haldane2008,
  title={Possible realization of directional optical waveguides in photonic crystals with broken time-reversal symmetry},
  author={Haldane, F. D. M. and Raghu, S.},
  journal={Phys. Rev. Lett.},
  volume={100},
  pages={013904},
  year={2008},
  doi={10.1103/PhysRevLett.100.013904},
  url={https://doi.org/10.1103/PhysRevLett.100.013904}
}

@article{Wang2009Nature,
  title={Observation of unidirectional backscattering-immune topological electromagnetic states},
  author={Wang, Zheng and Chong, Y. D. and Joannopoulos, J. D. and Soljačić, Marin},
  journal={Nature},
  volume={461},
  pages={772--775},
  year={2009},
  doi={10.1038/nature08293},
  url={https://doi.org/10.1038/nature08293}
}

@article{Poo2011,
  title = {Experimental Realization of Self-Guiding Unidirectional Electromagnetic Edge States},
  author = {Poo, Yin and Wu, Rui-xin and Lin, Zhifang and Yang, Yan and Chan, C. T.},
  journal = {Phys. Rev. Lett.},
  volume = {106},
  issue = {9},
  pages = {093903},
  numpages = {4},
  year = {2011},
  month = {Mar},
  publisher = {American Physical Society},
  doi = {10.1103/PhysRevLett.106.093903},
  url = {https://link.aps.org/doi/10.1103/PhysRevLett.106.093903}
}

@article{Lu2014,
  title={Topological photonics},
  author={Lu, Ling and Joannopoulos, J. D. and Soljačić, Marin},
  journal={Nat. Photonics},
  volume={8},
  pages={821--829},
  year={2014},
  doi={10.1038/nphoton.2014.248},
  url={https://doi.org/10.1038/nphoton.2014.248}
}

@article{Ozawa2019,
  title={Topological photonics},
  author={Ozawa, Tomoki and Price, Hannah M. and Amo, Alberto and Goldman, Nathan and Hafezi, Mohammad and Lu, Ling and Rechtsman, Mikael C. and Schuster, David and Simon, Jonathan and Zilberberg, Oded and Carusotto, Iacopo},
  journal={Rev. Mod. Phys.},
  volume={91},
  pages={015006},
  year={2019},
  doi={10.1103/RevModPhys.91.015006},
  url={https://doi.org/10.1103/RevModPhys.91.015006}
}

@article{khanikaev2024,
  author = {Khanikaev, Alexander B. and Al{\`u}, Andrea},
  title = {Topological photonics: robustness and beyond},
  journal = {Nat. Commun.},
  volume = {15},
  number = {1},
  pages = {931},
  year = {2024},
  doi = {10.1038/s41467-024-45194-2},
  url = {https://doi.org/10.1038/s41467-024-45194-2},
  issn = {2041-1723}
  }

@article{RuoYang2022,
  title = {Photonic ${\mathbb{Z}}_{2}$ Topological Anderson Insulators},
  author = {Cui, Xiaohan and Zhang, Ruo-Yang and Zhang, Zhao-Qing and Chan, C. T.},
  journal = {Phys. Rev. Lett.},
  volume = {129},
  issue = {4},
  pages = {043902},
  numpages = {7},
  year = {2022},
  month = {Jul},
  publisher = {American Physical Society},
  doi = {10.1103/PhysRevLett.129.043902},
  url = {https://link.aps.org/doi/10.1103/PhysRevLett.129.043902}
}

@article{Guglielmon2019,
  title = {Broadband Topological Slow Light through Higher Momentum-Space Winding},
  author = {Guglielmon, Jonathan and Rechtsman, Mikael C.},
  journal = {Phys. Rev. Lett.},
  volume = {122},
  issue = {15},
  pages = {153904},
  numpages = {5},
  year = {2019},
  month = {Apr},
  publisher = {American Physical Society},
  doi = {10.1103/PhysRevLett.122.153904},
  url = {https://link.aps.org/doi/10.1103/PhysRevLett.122.153904}
}

@article{yu2021topological,
    author = {Yu, Letian and Xue, Haoran and Zhang, Baile},
    title = {Topological slow light via coupling chiral edge modes with flatbands},
    journal = {Appl. Phys. Lett.},
    volume = {118},
    number = {7},
    pages = {071102},
    year = {2021},
    month = {02},
    issn = {0003-6951},
    doi = {10.1063/5.0039839},
    url = {https://doi.org/10.1063/5.0039839}
    }

@article{Rechtsman2013,
  title={Photonic Floquet topological insulators},
  author={Rechtsman, Mikael C. and Zeuner, Julia M. and Plotnik, Yonatan and Lumer, Yaakov and Podolsky, Daniel and Dreisow, Felix and Nolte, Stefan and Segev, Mordechai},
  journal={Nature},
  volume={496},
  pages={196--200},
  year={2013},
  doi={10.1038/nature12066},
  url={https://doi.org/10.1038/nature12066}
}

@article{Mann2021Broadband,
  title = {Broadband Topological Slow Light through Brillouin Zone Winding},
  author = {Mann, Sander A. and Al\`u, Andrea},
  journal = {Phys. Rev. Lett.},
  volume = {127},
  issue = {12},
  pages = {123601},
  numpages = {6},
  year = {2021},
  month = {Sep},
  publisher = {American Physical Society},
  doi = {10.1103/PhysRevLett.127.123601},
  url = {https://link.aps.org/doi/10.1103/PhysRevLett.127.123601}
}

@article{Ma2016Valley,
doi = {10.1088/1367-2630/18/2/025012},
url = {https://doi.org/10.1088/1367-2630/18/2/025012},
year = {2016},
month = {feb},
publisher = {IOP Publishing},
volume = {18},
number = {2},
pages = {025012},
author = {Ma, Tzuhsuan and Shvets, Gennady},
title = {All-Si valley-Hall photonic topological insulator},
journal = {New J. Phys.}
}

@article{Kumar2024,
  author   = {Abhishek Kumar and Yi Ji Tan and Nikhil Navaratna and Manoj Gupta and Prakash Pitchappa and Ranjan Singh},
  title    = {Slow light topological photonics with counter-propagating waves and its active control on a chip},
  journal  = {Nat. Commun.},
  year     = {2024},
  volume   = {15},
  number   = {1},
  pages    = {926},
  month    = {jan},
  doi      = {10.1038/s41467-024-45175-5}
}

@Article{Peng2025,
AUTHOR = {Peng, Chenyang and Li, Gang and Yang, Junhao and Ma, Chunlin and Qi, Xinyuan},
TITLE = {Tunable Slow Light in Valley-Locked Topological Photonic Crystal Waveguide},
JOURNAL = {Photonics},
VOLUME = {12},
YEAR = {2025},
NUMBER = {4},
pages = {332},
URL = {https://www.mdpi.com/2304-6732/12/4/332},
DOI = {https://doi.org/10.3390/photonics12040332}
}

@article{Kitagawa2010Floquet,
  title={Topological characterization of periodically driven quantum systems},
  author={Kitagawa, T. and Berg, E. and Rudner, M. and Demler, E.},
  journal={Phys. Rev. B},
  volume={82},
  pages={235114},
  year={2010},
  doi={10.1103/PhysRevB.82.235114},
  url={https://doi.org/10.1103/PhysRevB.82.235114}
}

@article{Hafezi2013,
  title={Imaging topological edge states in silicon photonics},
  author={Hafezi, M. and Mittal, S. and Fan, J. and Migdall, A. and Taylor, J.},
  journal={Nat. Photonics},
  volume={7},
  pages={1001--1005},
  year={2013},
  doi={10.1038/nphoton.2013.274},
  url={https://doi.org/10.1038/nphoton.2013.274}
}

@article{Wang2025TESC,
  author = {Wang, Wenhao and Shen, Zhonglei and Tan, Yi Ji and Chen, Kaiji and Singh, Ranjan},
  title = {On-chip topological edge state cavities},
  journal = {Light Sci. Appl.},
  volume = {14},
  number = {1},
  pages = {330},
  year = {2025},
  doi = {10.1038/s41377-025-02017-3},
  url = {https://doi.org/10.1038/s41377-025-02017-3},
  issn = {2047-7538}
  }

@article{Harper1955,
doi = {10.1088/0370-1298/68/10/304},
url = {https://doi.org/10.1088/0370-1298/68/10/304},
year = {1955},
month = {oct},
publisher = {},
volume = {68},
number = {10},
pages = {874},
author = {P G Harper},
title = {Single Band Motion of Conduction Electrons in a Uniform Magnetic Field},
journal = {Proc. Phys. Soc. A}
}

@article{Hofstadter1976,
  title = {Energy levels and wave functions of Bloch electrons in rational and irrational magnetic fields},
  author = {Hofstadter, Douglas R.},
  journal = {Phys. Rev. B},
  volume = {14},
  issue = {6},
  pages = {2239--2249},
  numpages = {0},
  year = {1976},
  month = {Sep},
  publisher = {American Physical Society},
  doi = {10.1103/PhysRevB.14.2239},
  url = {https://link.aps.org/doi/10.1103/PhysRevB.14.2239}
}

@article{Harper2014,
  title = {Perturbative approach to flat Chern bands in the Hofstadter model},
  author = {Harper, Fenner and Simon, Steven H. and Roy, Rahul},
  journal = {Phys. Rev. B},
  volume = {90},
  issue = {7},
  pages = {075104},
  numpages = {36},
  year = {2014},
  month = {Aug},
  publisher = {American Physical Society},
  doi = {10.1103/PhysRevB.90.075104},
  url = {https://link.aps.org/doi/10.1103/PhysRevB.90.075104}
}

@article{Gunnar2015,
  title = {Fractional Chern Insulators in Harper-Hofstadter Bands with Higher Chern Number},
  author = {M\"oller, Gunnar and Cooper, Nigel R.},
  journal = {Phys. Rev. Lett.},
  volume = {115},
  issue = {12},
  pages = {126401},
  numpages = {6},
  year = {2015},
  month = {Sep},
  publisher = {American Physical Society},
  doi = {10.1103/PhysRevLett.115.126401},
  url = {https://link.aps.org/doi/10.1103/PhysRevLett.115.126401}
}

@article{Wauters2018,
  title = {Quantization of the Hall conductivity in the Harper-Hofstadter model},
  author = {Wauters, Matteo M. and Santoro, Giuseppe E.},
  journal = {Phys. Rev. B},
  volume = {98},
  issue = {20},
  pages = {205112},
  numpages = {13},
  year = {2018},
  month = {Nov},
  publisher = {American Physical Society},
  doi = {10.1103/PhysRevB.98.205112},
  url = {https://link.aps.org/doi/10.1103/PhysRevB.98.205112}
}

@article{Zhong2023,
  author   = {Zhong-Cheng Xiang and Kaixuan Huang and Yu-Ran Zhang and Tao Liu and Yun-Hao Shi and Cheng-Lin Deng and Tong Liu and Hao Li and Gui-Han Liang and Zheng-Yang Mei and Haifeng Yu and Guangming Xue and Ye Tian and Xiaohui Song and Zhi-Bo Liu and Kai Xu and Dongning Zheng and Franco Nori and Heng Fan},
  title    = {Simulating Chern insulators on a superconducting quantum processor},
  journal  = {Nat. Commun.},
  year     = {2023},
  volume   = {14},
  number   = {1},
  pages    = {5433},
  month    = {sep},
  doi      = {10.1038/s41467-023-41230-9}
}

@article{Ye:22,
author = {Fan Ye and Xiankai Sun},
journal = {Opt. Express},
number = {15},
pages = {26620--26627},
publisher = {Optica Publishing Group},
title = {Hofstadter butterfly and topological edge states in a quasiperiodic photonic crystal cavity array},
volume = {30},
month = {Jul},
year = {2022},
url = {https://opg.optica.org/oe/abstract.cfm?URI=oe-30-15-26620},
doi = {10.1364/OE.453985}
}

@article{Xue2023,
  author   = {Xue Han and Fude Li and De-Xiu Qiu and Kang Xue and X. X. Yi},
  title    = {Quantized fields induced topological features in Harper-Hofstadter model},
  journal  = {AAPPS Bulletin},
  year     = {2023},
  volume   = {33},
  number   = {1},
  pages    = {1},
  month    = {jan},
  issn     = {2309-4710},
  doi      = {10.1007/s43673-022-00071-2}
}

@article{DUTRA2023114338,
title = {Hofstadter butterfly in optical multilayers},
journal = {Opt. Mater.},
volume = {144},
pages = {114338},
year = {2023},
issn = {0925-3467},
doi = {https://doi.org/10.1016/j.optmat.2023.114338},
url = {https://www.sciencedirect.com/science/article/pii/S0925346723009102},
author = {R.F. Dutra and M.S. Vasconcelos and D.H.A.L. Anselmo}
}

@article{Lin2022All,
  title = {All-Optical Control of the Photonic Hall Lattice in a Pumped Waveguide Array},
  author = {Lin, Shirong and Wang, Luojia and Yuan, Luqi and Chen, Xianfeng},
  journal = {Phys. Rev. Appl},
  volume = {17},
  issue = {6},
  pages = {064029},
  numpages = {10},
  year = {2022},
  month = {Jun},
  publisher = {American Physical Society},
  doi = {10.1103/PhysRevApplied.17.064029},
  url = {https://link.aps.org/doi/10.1103/PhysRevApplied.17.064029}
}

@article{Matteo2019,
  title = {Theory of chiral edge state lasing in a two-dimensional topological system},
  author = {Secl\`{\i}, Matteo and Capone, Massimo and Carusotto, Iacopo},
  journal = {Phys. Rev. Res.},
  volume = {1},
  issue = {3},
  pages = {033148},
  numpages = {10},
  year = {2019},
  month = {Dec},
  publisher = {American Physical Society},
  doi = {10.1103/PhysRevResearch.1.033148},
  url = {https://link.aps.org/doi/10.1103/PhysRevResearch.1.033148}
}

@article{Amelio2020,
  title = {Theory of the Coherence of Topological Lasers},
  author = {Amelio, Ivan and Carusotto, Iacopo},
  journal = {Phys. Rev. X},
  volume = {10},
  issue = {4},
  pages = {041060},
  numpages = {20},
  year = {2020},
  month = {Dec},
  publisher = {American Physical Society},
  doi = {10.1103/PhysRevX.10.041060},
  url = {https://link.aps.org/doi/10.1103/PhysRevX.10.041060}
}

@article{Tai2017,
  title   = {Microscopy of the interacting Harper--Hofstadter model in the two-body limit},
  author  = {Tai, M. Eric and Lukin, Alexander and Rispoli, Matthew and Schittko, Robert and Menke, Tim and Borgnia, Dan and Preiss, Philipp M. and Grusdt, Fabian and Kaufman, Adam M. and Greiner, Markus},
  journal = {Nature},
  volume  = {546},
  number  = {7659},
  pages   = {519--523},
  year    = {2017},
  doi     = {10.1038/nature22811}
}

@article{Vega2023,
  title = {Topological multimode waveguide {QED}},
  author = {Vega, C. and Porras, D. and Gonz\'alez-Tudela, A.},
  journal = {Phys. Rev. Res.},
  volume = {5},
  issue = {2},
  pages = {023031},
  numpages = {19},
  year = {2023},
  month = {Apr},
  publisher = {American Physical Society},
  doi = {10.1103/PhysRevResearch.5.023031},
  url = {https://link.aps.org/doi/10.1103/PhysRevResearch.5.023031}
}

@article{Gyunghun2024,
  author   = {Gyunghun Kim and Joseph Suh and Dayeong Lee and Namkyoo Park and Sunkyu Yu},
  title    = {Long-range-interacting topological photonic lattices breaking channel-bandwidth limit},
  journal  = {Light Sci. Appl.},
  year     = {2024},
  volume   = {13},
  pages    = {189},
  month    = {sep},
  doi      = {10.1038/s41377-024-01557-4}
}

@article{Jia-Qi2023,
  title = {Light-matter interactions in a Hofstadter lattice with next-nearest-neighbor couplings},
  author = {Li, Jia-Qi and Gao, Zhao-Min and Liu, Wen-Xiao and Wang, Xin},
  journal = {Phys. Rev. A},
  volume = {108},
  issue = {4},
  pages = {043708},
  numpages = {10},
  year = {2023},
  month = {Oct},
  publisher = {American Physical Society},
  doi = {10.1103/PhysRevA.108.043708},
  url = {https://link.aps.org/doi/10.1103/PhysRevA.108.043708}
}

@article{Wu2020Nontrivial,
  title = {Nontrivial topological phase with a zero Chern number},
  author = {Wu, H. C. and Jin, L. and Song, Z.},
  journal = {Phys. Rev. B},
  volume = {102},
  issue = {3},
  pages = {035145},
  numpages = {8},
  year = {2020},
  month = {Jul},
  publisher = {American Physical Society},
  doi = {10.1103/PhysRevB.102.035145},
  url = {https://link.aps.org/doi/10.1103/PhysRevB.102.035145}
}

@article{Leykam2018,
  title = {Reconfigurable Topological Phases in Next-Nearest-Neighbor Coupled Resonator Lattices},
  author = {Leykam, Daniel and Mittal, S. and Hafezi, M. and Chong, Y. D.},
  journal = {Phys. Rev. Lett.},
  volume = {121},
  issue = {2},
  pages = {023901},
  numpages = {6},
  year = {2018},
  month = {Jul},
  publisher = {American Physical Society},
  doi = {10.1103/PhysRevLett.121.023901},
  url = {https://link.aps.org/doi/10.1103/PhysRevLett.121.023901}
}

@article{Bell:17,
author = {Bryn A. Bell and Kai Wang and Alexander S. Solntsev and Dragomir N. Neshev and Andrey A. Sukhorukov and Benjamin J. Eggleton},
journal = {Optica},
number = {11},
pages = {1433--1436},
publisher = {Optica Publishing Group},
title = {Spectral photonic lattices with complex long-range coupling},
volume = {4},
month = {Nov},
year = {2017},
url = {https://opg.optica.org/optica/abstract.cfm?URI=optica-4-11-1433},
doi = {10.1364/OPTICA.4.001433}
}

@article{Pellerin2024,
  title = {Wave-Function Tomography of Topological Dimer Chains with Long-Range Couplings},
  author = {Pellerin, F. and Houvenaghel, R. and Coish, W. A. and Carusotto, I. and St-Jean, P.},
  journal = {Phys. Rev. Lett.},
  volume = {132},
  issue = {18},
  pages = {183802},
  numpages = {6},
  year = {2024},
  month = {May},
  publisher = {American Physical Society},
  doi = {10.1103/PhysRevLett.132.183802},
  url = {https://link.aps.org/doi/10.1103/PhysRevLett.132.183802}
}

@article{Wang2024Unconvent,
author = {Wang, Pengfei and Huang, Lei and Zhang, Hanxiao and Yang, Hong and Yan, Dong},
title = {Unconventional Light-Matter Interactions Between Giant Atoms and Structured Baths with Next-Nearest-Neighbor Couplings},
journal = {Annalen der Physik},
volume = {536},
number = {10},
pages = {2400165},
doi = {https://doi.org/10.1002/andp.202400165},
url = {https://onlinelibrary.wiley.com/doi/abs/10.1002/andp.202400165},
year = {2024}
}

@article{Chuan2021,
author = {Chuan-Xun Du and Nan Xu and Lei Du and Yan Zhang and Jin-Hui Wu},
journal = {Opt. Express},
number = {23},
pages = {37722--37732},
publisher = {Optica Publishing Group},
title = {Topological edge states controlled by next-nearest-neighbor coupling and Peierls phase in a P            T-symmetric trimerized lattice},
volume = {29},
month = {Nov},
year = {2021},
url = {https://opg.optica.org/oe/abstract.cfm?URI=oe-29-23-37722},
doi = {10.1364/OE.438779}
}

@article{keil2015direct,
  title={Direct measurement of second-order coupling in a waveguide lattice},
  author={Keil, Robert and Pressl, Benedikt and Heilmann, Ren{\'e} and Gr{\"a}fe, Markus and Weihs, Gregor and Szameit, Alexander},
  journal={Appl. Phys. Lett.},
  volume={107},
  number={24},
  year={2015},
  publisher={AIP Publishing},
  url = {https://doi.org/10.1063/1.4937807}
}

@article{Hasan2010,
  title = {Colloquium: Topological insulators},
  author = {Hasan, M. Z. and Kane, C. L.},
  journal = {Rev. Mod. Phys.},
  volume = {82},
  issue = {4},
  pages = {3045--3067},
  numpages = {0},
  year = {2010},
  month = {Nov},
  publisher = {American Physical Society},
  doi = {10.1103/RevModPhys.82.3045},
  url = {https://link.aps.org/doi/10.1103/RevModPhys.82.3045}
}

@article{Pertsch2002,
  title = {Anomalous Refraction and Diffraction in Discrete Optical Systems},
  author = {Pertsch, T. and Zentgraf, T. and Peschel, U. and Br\"auer, A. and Lederer, F.},
  journal = {Phys. Rev. Lett.},
  volume = {88},
  issue = {9},
  pages = {093901},
  numpages = {4},
  year = {2002},
  month = {Feb},
  publisher = {American Physical Society},
  doi = {10.1103/PhysRevLett.88.093901},
  url = {https://link.aps.org/doi/10.1103/PhysRevLett.88.093901}
}

@article{Regulating2025,
author = {Qin, Chengzhi and Liu, Menglin and Wang, Bing and Longhi, Stefano and Lu, Peixiang},
title = {Regulating Light Refraction and Reflection Using Speed-Tailored Optical Potentials},
journal = {Laser Photonics Rev.},
volume = {19},
number = {20},
pages = {e00206},
doi = {https://doi.org/10.1002/lpor.202500206},
url = {https://onlinelibrary.wiley.com/doi/abs/10.1002/lpor.202500206},
year = {2025}
}

@article{Maczewsky2017,
  author = {Maczewsky, Lukas J. and Zeuner, Julia M. and Nolte, Stefan and Szameit, Alexander},
  title = {Observation of photonic anomalous Floquet topological insulators},
  journal = {Nat. Commun.},
  year = {2017},
  volume = {8},
  number = {1},
  pages = {13756},
  doi = {10.1038/ncomms13756},
  url = {https://doi.org/10.1038/ncomms13756}
  }

@article{Sebabrata2020,
author = {Sebabrata Mukherjee  and Mikael C. Rechtsman },
title = {Observation of Floquet solitons in a topological bandgap},
journal = {Science},
volume = {368},
number = {6493},
pages = {856-859},
year = {2020},
doi = {10.1126/science.aba8725},
URL = {https://www.science.org/doi/abs/10.1126/science.aba8725}
}

@article{Szameit_2010,
doi = {10.1088/0953-4075/43/16/163001},
url = {https://doi.org/10.1088/0953-4075/43/16/163001},
year = {2010},
month = {jul},
publisher = {},
volume = {43},
number = {16},
pages = {163001},
author = {Szameit, Alexander and Nolte, Stefan},
title = {Discrete optics in femtosecond-laser-written photonic structures},
journal = {J. Phys. B: At. Mol. Opt. Phys.}
}

@article{Mrejen2015,
  author = {Mrejen, Michael and Suchowski, Haim and Hatakeyama, Taiki and Wu, Chihhui and Feng, Liang and O'Brien, Kevin and Wang, Yuan and Zhang, Xiang},
  title = {Adiabatic elimination-based coupling control in densely packed subwavelength waveguides},
  journal = {Nat. Commun.},
  year = {2015},
  volume = {6},
  number = {1},
  pages = {7565},
  date = {2015-06-26},
  doi = {10.1038/ncomms8565},
  url = {https://doi.org/10.1038/ncomms8565}
  }

\end{document}